\documentclass[twocolumn,aps,superscriptaddress,showpacs,amsmath,amssymb,prl]{revtex4}
\usepackage{graphicx}
\usepackage[T1]{fontenc}
\usepackage{currvita}
\usepackage{units}
\usepackage{xspace}
\usepackage{color}
\usepackage{soul} 
\usepackage[normalem]{ulem} 

\begin{document}

\title{Exchange interaction and its tuning in magnetic binary chalcogenides}

\author{M.\,G. Vergniory}
\email{mgarcia@mpi-halle.de}
\affiliation{Max-Planck-Institut f\"ur Mikrostrukturphysik, Weinberg 2, 06120 Halle, Germany}
\author{D. Thonig}
\affiliation{Max-Planck-Institut f\"ur Mikrostrukturphysik, Weinberg 2, 06120 Halle, Germany}
\author{M. Hoffmann}
\affiliation{Max-Planck-Institut f\"ur Mikrostrukturphysik, Weinberg 2, 06120 Halle, Germany}.  
\affiliation{Institut f\"ur Physik, Martin-Luther-Universit\"at Halle-Wittenberg, D-06099 Halle, Germany}
\author{I.\,V. Maznichenko}
\affiliation{Institut f\"ur Physik, Martin-Luther-Universit\"at Halle-Wittenberg, D-06099 Halle, Germany}
\author{M. Geilhufe}
\affiliation{Max-Planck-Institut f\"ur Mikrostrukturphysik, Weinberg 2, 06120 Halle, Germany}
\author{M.\,M. Otrokov}
\affiliation{Physics Department, Tomsk State University, pr. Lenina 36, 634050 Tomsk, Russia}
\author{X.\, Zubizarreta}
\affiliation{Max-Planck-Institut f\"ur Mikrostrukturphysik, Weinberg 2, 06120 Halle, Germany}
\affiliation{Donostia International Physics Center, P. de Manuel Lardizabal 4, San Sebasti\'an, 20018 Basque Country, Spain}
\author{S. Ostanin}
\affiliation{Max-Planck-Institut f\"ur Mikrostrukturphysik, Weinberg 2, 06120 Halle, Germany}
\author{A. Marmodoro}
\affiliation{Max-Planck-Institut f\"ur Mikrostrukturphysik, Weinberg 2, 06120 Halle, Germany}
\author{J. Henk}
\affiliation{Institut f\"ur Physik, Martin-Luther-Universit\"at Halle-Wittenberg, D-06099 Halle, Germany}
\author{W. Hergert}
\affiliation{Institut f\"ur Physik, Martin-Luther-Universit\"at Halle-Wittenberg, D-06099 Halle, Germany}
\author{I. Mertig}
\affiliation{Max-Planck-Institut f\"ur Mikrostrukturphysik, Weinberg 2, 06120 Halle, Germany}
\affiliation{Institut f\"ur Physik, Martin-Luther-Universit\"at Halle-Wittenberg, D-06099 Halle, Germany}
\author{E.\,V. Chulkov}
\affiliation{Physics Department, Tomsk State University, pr. Lenina 36, 634050 Tomsk, Russia}
\affiliation{Donostia International Physics Center, P. de Manuel Lardizabal 4, San Sebasti\'an, 20018 Basque Country, Spain}
\affiliation{Departamento  de F\'isica de Materiales, Facultad de Ciencias Qu\'imicas, Apdo. 1072, San Sebasti\'an, 20080 Basque Country, Spain}
\author{A. Ernst}
\email{aernst@mpi-halle.de}
\affiliation{Max-Planck-Institut f\"ur Mikrostrukturphysik, Weinberg 2, 06120 Halle, Germany}

\begin{abstract}
  Using a first-principles Green's function approach we study magnetic
  properties of the magnetic binary chalcogenides Bi$_{2}$Te$_{3}$,
  Bi$_{2}$Se$_{3}$, and Sb$_{2}$Te$_{3}$. The magnetic coupling
  between transition-metal impurities is long-range, extends beyond a
  quintuple layer, and decreases with increasing number of $d$
  electrons per $3d$ atom. We find two main mechanisms for the
  magnetic interaction in these materials: the indirect exchange
  interaction mediated by free carriers and the indirect interaction
  between magnetic moments via chalcogen atoms.  The calculated Curie
  temperatures of these systems are in good agreement with available
  experimental data. Our results provide deep insight into magnetic
  interactions in magnetic binary chalcogenides and open a way to
  design new materials for promising applications.
\end{abstract}

\pacs{71.70.Ej,  72.15.Jf, 85.80.Fi}

\date{\today }

\maketitle
Tetradymite chalcogenides, in particular Bi$_2$Te$_3$, Bi$_2$Se$_3$,
and Sb$_2$Te$_3$, are of great interest due to their outstanding
structural and electronic properties. These compounds consist of
repeated blocks of five atomic layers (quintuple layers) separated by
a van der Waals gap. The electronic structure features a narrow band
gap and strong spin-orbit coupling, which are responsible for the
inverted band structure at the Brillouin zone center $\Gamma$.
Tetradymite chalcogenides are attractive for thermoelectric
applications \cite{Nolas2001} because of their high figure of merit at
room temperature.  Recently, topologically protected surface states
have been observed in all of these chalcogenides, which makes them
subject of intense research \cite{Hasan2010,Qi2011}.  Especially,
these bichalcogenides serve as a basis for new materials with desired
properties \cite{Eremeev2012}. This is feasible by stacking of
different compounds or specific doping. In particular, doping with
magnetic impurities can open new perspectives for spintronics and spin
caloritronics applications
\cite{Zutic2004,Sato2010,Uchida2008,Bauer2012}.

The aim of this work is to study the magnetic properties of
tetradymite chalcogenides doped with transition metal impurities. Some
of these systems were already studied as possible candidates for
spintronics applications some years ago
\cite{Dyck2002,Dyck2003,Dyck2005,Kulbachinskii2003,Kulbachinskii2005,Choi2004,Zhou2005,Zhou2006a,Zhou2006c,Bos2006,Hor2010}.
Most of the experiments were done on single crystals with a maximal
doping concentration of $x=0.1$. Thereby, stable ferromagnetic order was
observed mostly in chalcogenides doped with vanadium and chromium
\cite{Dyck2002,Dyck2005,Kulbachinskii2005}, while samples doped with
manganese were found either ferromagnetic at very low temperatures or
antiferro- and paramagnetic depending on experimental conditions and sample
preparation
\cite{Dyck2003,Kulbachinskii2003,Zhou2006c,Bos2006,Hor2010}.  High
Curie temperatures were reported for films prepared with
molecular-beam epitaxy: $177$ K and $190K$ K for
Sb$_{1.65}$V$_{0.35}$Te$_3$ and Sb$_{1.41}$Cr$_{0.59}$Te$_3$,
respectively \cite{Zhou2005,Zhou2006a}. The great interest in breaking of time reversal symmetry in
topological insulators motivated further investigations of magnetic
impurities located in particular at the surfaces of tetradymite
chalcogenides
\cite{Hsieh2009b,Hor2010,Honolka2012,Shelford2012,West2012,Scholz2012,Kou2012,Ye2012}.

Magnetic properties of magnetic chalcogenides can be
efficiently described by first-principles methods.
One of the first comprehensive studies was carried out by Larson and
Lambrecht \cite{Larson2008}, who investigated the electronic and
magnetic properties of bulk Bi$_2$Te$_3$, Bi$_2$Se$_3$, and
Sb$_2$Te$_3$ doped with $3d$ transition metal atoms; their results
for magnetically doped Bi$_2$Se$_3$ were confirmed by several groups
\cite{Yu2010,Zhang2012}. Recently, it was shown that the Dirac
surface state of the topological insulator Bi$_2$Te$_3$ survives upon
moderate Mn doping of the surface layer, but can lose its topological
nontrivial character depending on the magnetization direction
\cite{Henk2012,Henk2012a}. However, critical magnetic properties and
the exchange interaction in magnetic chalcogenides were not
studied in detail on a theoretical \textit{ab initio} level and, thus,
are still under debate.

In this work, doping bulk tetradymite chalcogenides with transition
metals by means of a first-principles Green's function method, we show
that the exchange interaction in these materials can be either long-range
ferromagnetic, antiferromagnetic or paramagnetic, depending on the
host and the impurity atoms. We identify in particular two main types of magnetic
interactions and discuss ways to manipulate the magnetic properties of
these systems.

The calculations were performed within the density functional theory
using the local spin density approximation (LSDA) and the generalized gradient
approximation (GGA) \cite{Perdew1992a,Perdew1996}. A self-consistent Green's
function method in both relativistic and scalar-relativistic
implementations was used to compute electronic and magnetic properties
of the magnetic chalcogenides. Substitutional disorder was treated
within the coherent potential approximation (CPA; e.\,g.\
\cite{Gyorffy1972}). Heisenberg exchange constants $J_{ij}$ were
obtained using the magnetic force theorem as it is implemented within the
multiple-scattering theory \cite{Liechtenstein1987}. Inclusion of
spin-orbit coupling leads to minor changes in the magnetic interaction
(about 3-5\% with respect to the scalar-relativistic case). Therefore,
for the sake of clarity, here we present only exchange constants
calculated within the scalar-relativistic approximation.

According to the available experimental data
\cite{Dyck2002,Dyck2003,Dyck2005,Kulbachinskii2003,Kulbachinskii2005,
  Choi2004,Zhou2005,Zhou2006a,Zhou2006c,Bos2006,Hor2010,Hsieh2009b,Scholz2012,Kou2012},
3$d$ transition metal impurities in bulk tetradymite chalcogenides
substitute typically cation atoms (Bi and Sb) and can supply 1--3
electrons for the bonding. The comparably smaller size of transition
metal ions may lead to substantial relaxations of the underlying
crystal structure \cite{Larson2008}. We did not account for such
structural deformations in our CPA calculations but investigated their
impact on the magnetic interaction using a supercell approach, and
found only minor changes of the exchange constant values. Therefore,
the discussion below reports results from CPA calculations.

\begin{figure}[htb]
  \centering
  \includegraphics[width=0.85\columnwidth]{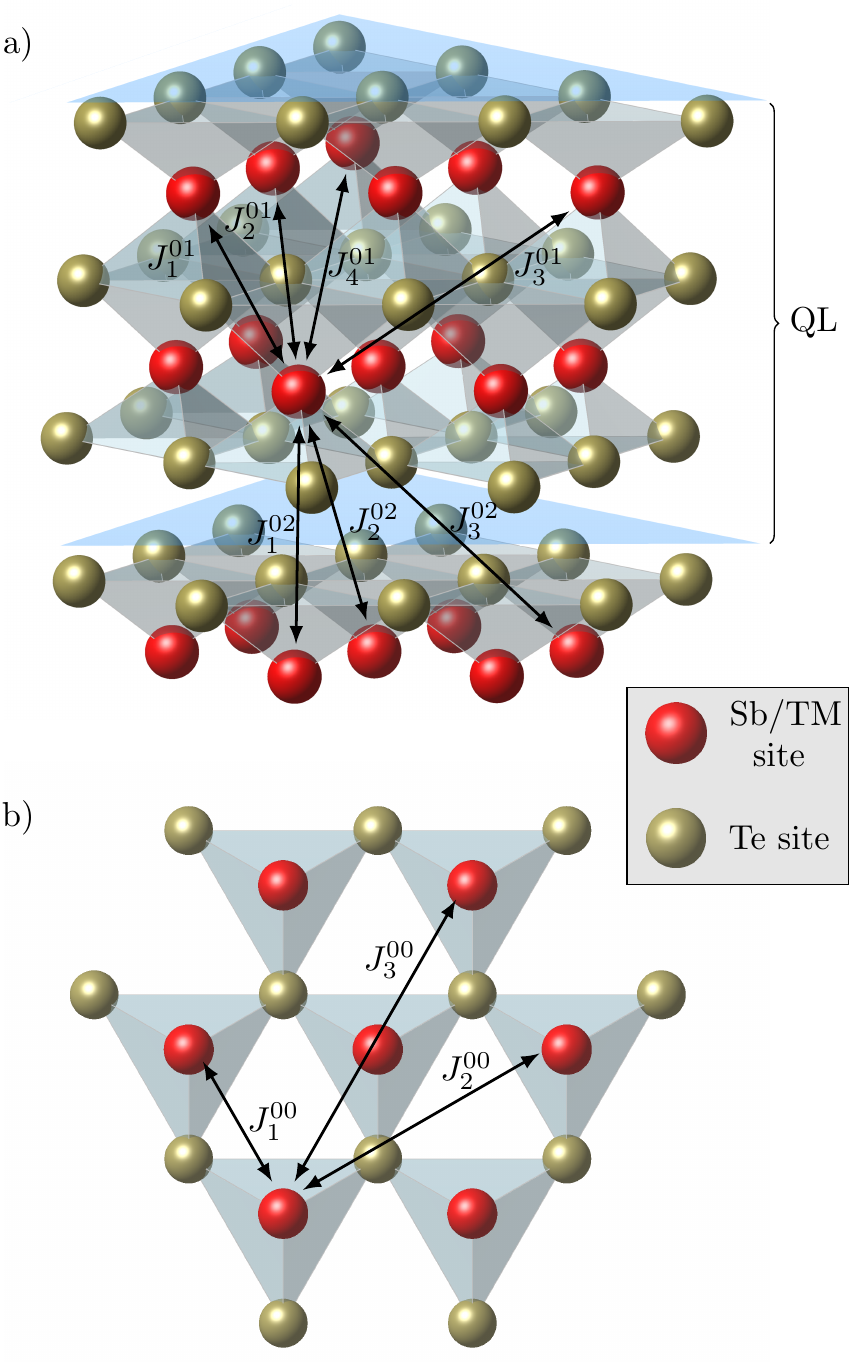}
  \caption{\label{fig:1} (Color online) Schematic view of magnetic
    interactions in magnetic Sb$_{2-x}$TM$_{x}$Te$_3$: exchange
    interactions between different layers (a) and within a single
    layer (b), respectively.  The corresponding exchange constants for
    Ti, V, Cr, Mn, Fe, and Co are shown in Fig.~\ref{fig:2}. The same
    plot applies to Bi$_{2-x}$TM$_x$Te$_3$ and Bi$_{2-x}$TM$_x$Se$_3$.}
\end{figure}

\begin{figure}[htb]
  \centering
  \includegraphics[width=\columnwidth]{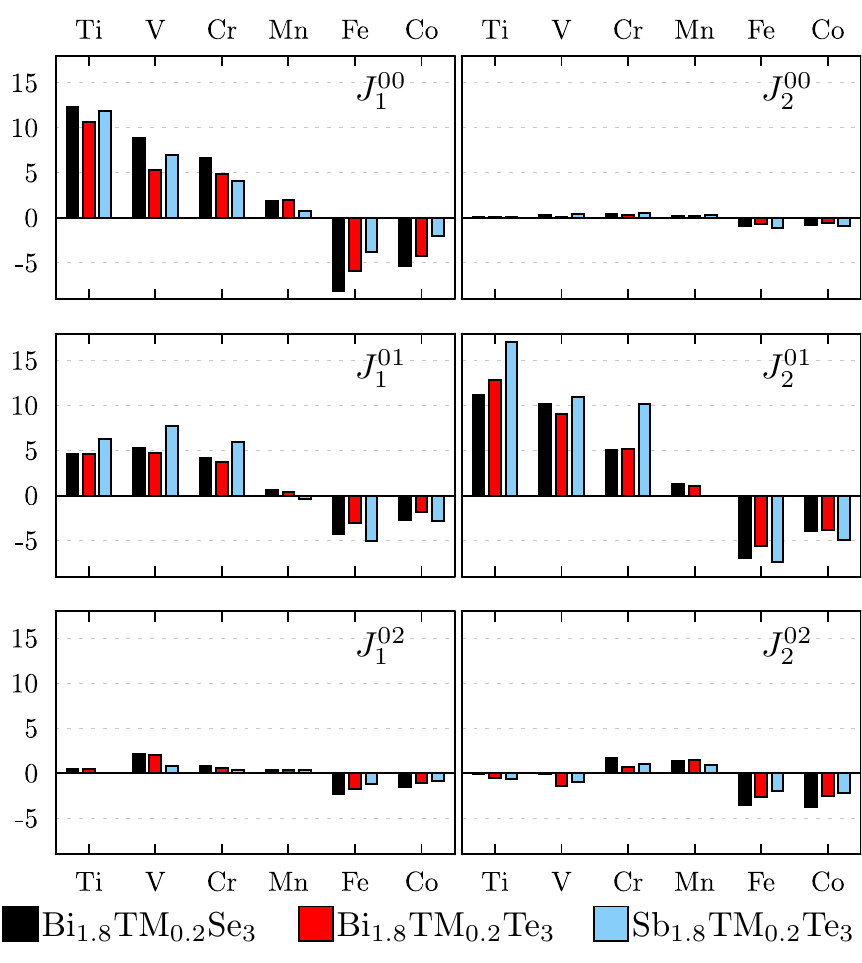}
  \caption{\label{fig:2}(Color online) Exchange constants (in meV) in 
    magnetic Bi$_{1.80}$TM$_{0.20}$Se$_3$,
    Bi$_{1.80}$TM$_{0.20}$Te$_3$, and Sb$_{1.80}$TM$_{0.20}$Te$_3$ (TM
    = Ti, V, Cr, Mn, Fe, Co). See the schematic view of magnetic
    interactions in Fig.~\ref{fig:1}.}
\end{figure}

We performed extensive Green's function calculations of
Bi$_{2-x}$TM$_x$Se$_3$, Bi$_{2-x}$TM$_x$Te$_3$, and
Sb$_{2-x}$TM$_x$Te$_3$ (TM = Ti, V, Cr, Mn, Fe, Co, Ni) for the range of
concentrations $0< x < 1.0$. The electronic structures of these
compounds calculated within the CPA agree for low and medium
concentrations ($x<0.3$) with those of previous supercell calculations
by Larson and Lambrecht \cite{Larson2008} (see the Supplementary Material). 
The self-consistently obtained Green's function was
further used to calculate the magnetic exchange constants $J_{ij}$. Their relevant directions are depicted in Fig.~\ref{fig:1} on top of the lattice structure of Sb$_{2-x}$TM$_x$Te$_3$ for clarity, where we distinguish among in-plane (within the Sb
or Bi plane) and an out-of-plane coupling. Although experimental data are
available for a large concentration range \cite{Zhou2005,Zhou2006a},
we first focus discussion on a representative value of $x =
0.2$. The results presented in Fig.~\ref{fig:2} can be summarized as
follows.

(i) The effective exchange interaction is reduced with increasing
number of $d$ electrons per TM, going from positive to negative
values. The strongest ferromagnetic interaction is found between Ti
atoms, which is explained by the local density of states (DOS) of the
impurities (shown in the Supplementary Material).  The $d$ electrons
of Ti atoms in the majority spin channel hybridize strongly with $sp$
states of the host.  The corresponding DOS shows a strong dispersion
(see the Supplementary Material) and is located mostly around the
Fermi level with one occupied $d$ orbital, while the $d$ bonds in
the minority spin channel are all unoccupied.  Thus, the net
magnetization is 1.0 $\mu_B$/atom, indicating a valency of
$3+$ for each Ti atom.  The strongly dispersed DOS at the Fermi energy
leads to a large ferromagnetic coupling between the nearest magnetic
moments within the Sb plane (Fig.~\ref{fig:2}).

(ii) With increasing number of $d$ electrons per TM atom, the spectral weight at
the Fermi level is reduced. This leads to a strong decrease of the exchange interactions in
Sb$_{2-x}$Mn$_{x}$Te$_3$, in which the majority and minority spin $d$
electrons are well separated in energy and show minor dispersion.  In
the case of Fe impurities, due to an occupied minority spin $d$ state
at the Fermi level, the magnitude of the exchange interaction
increases but becomes negative because of the large exchange splitting
and the isolated impurity-like character of the occupied $d$
orbitals. For Co and Ni impurities, the valency changes from $3+$ to
$2+$ and $1+$, respectively, reducing, thereby, the magnitude of the
exchange interaction, which remains negative. The exchange interaction
of Sb$_{2-x}$Ni$_{x}$Te$_3$, Bi$_{2-x}$Ni$_{x}$Te$_3$, and
Bi$_{2-x}$Ni$_{x}$Se$_3$ is very weak and is not discussed here.

(iii) For almost all cases, the strongest exchange interaction is found
between magnetic moments located in different Sb (Bi) planes but
within the same quintuple layer (panel $J_2^{01}$ in Figs.~\ref{fig:1}
and~\ref{fig:2}). The coupling weakens systematically with the number
of $d$ electrons. This interaction occurs via a Te (Se) atom lying
between two impurities and is of double exchange type. In addition,
the magnitude of $J_1^{01}$ is as large as for the in-plane interaction
between the nearest magnetic moments, which is an indirect exchange
interaction mediated by free carrier $sp$ states \cite{Zener1951}.
We thus conclude that two different exchange mechanisms, the double exchange
interaction via an anion and the indirect exchange coupling via free
carriers, determine the magnetic order in the TM doped
chalcogenides.

(iv) The size of cation atoms is crucial for the exchange
interaction.  The large size of atoms and, thus, the more spatially
extended wave functions, can lead to a strong hybridization with the
electronic states of the neighboring atoms. On the one hand, this can
increase the number of free carriers within the cation layer, favoring
the indirect exchange of Zener type \cite{Zener1951}.  On the other
hand, the strong binding between a cation (e.\,g.\ Bi) and an anion
(e.\,g.\ Te) reduces the number of valence electrons of the anion and,
thereby, reduces the strength of the double exchange interaction.
Therefore, in the case of Sb$_2$Te$_3$, the double exchange
interaction via Te atoms is significantly larger than that in
Bi$_2$Te$_3$ and Bi$_2$Se$_3$.
 
\begin{figure}[htb]
  \centering
  \includegraphics[width=0.85\columnwidth]{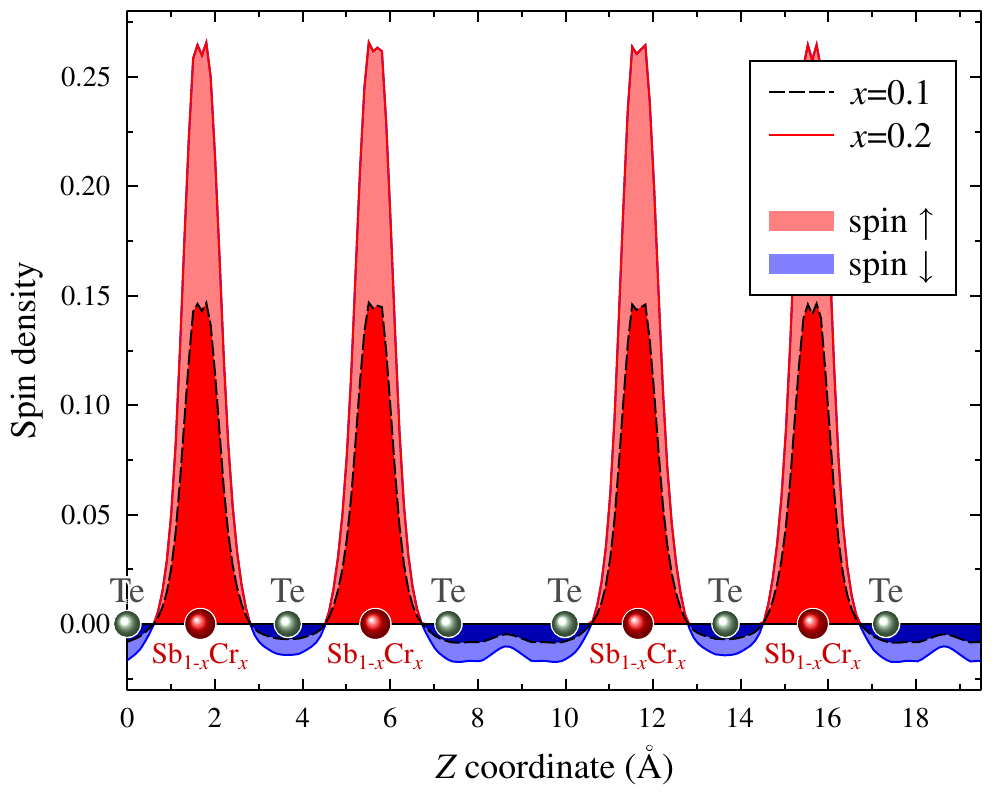}
  \caption{\label{fig:3}(color online) Spin density $m(z)$ (see the
    text) of Sb$_{2-x}$Cr$_{x}$Te$_3$ for $x = 0.1$ and $x = 0.2$ in
    the [0001] direction integrated over all in-plane coordinates $x$
    and $y$. The $z$ range covers two quintuple layers.}
\end{figure}

(v) Surprisingly the exchange interaction between magnetic moments located in
neighboring quintuple layers does not vanish (see $J_1^{02}$ and
$J_2^{02}$ in Figs.~\ref{fig:1} and~\ref{fig:2}). The spin
density $m(z) \equiv [\rho_\uparrow(z) - \rho_\downarrow(z)] /
[\rho_\uparrow(z) + \rho_\downarrow(z)]$ (where $\rho_\uparrow(z)$ and
$\rho_\downarrow(z)$ stand for the spin-up and -down charge densities,
respectively, integrated over the lateral coordinates $x$ and $y$)
``bridges'' the van der Waals gap and is responsible for the
``interquintuple layer'' magnetic interaction (Fig.~\ref{fig:3}). The
spin density in anion layers is negative and has a magnitude
comparable with that of the spin density in the van der Waals gap.

\begin{figure}
\centering
\includegraphics[width=0.90\columnwidth]{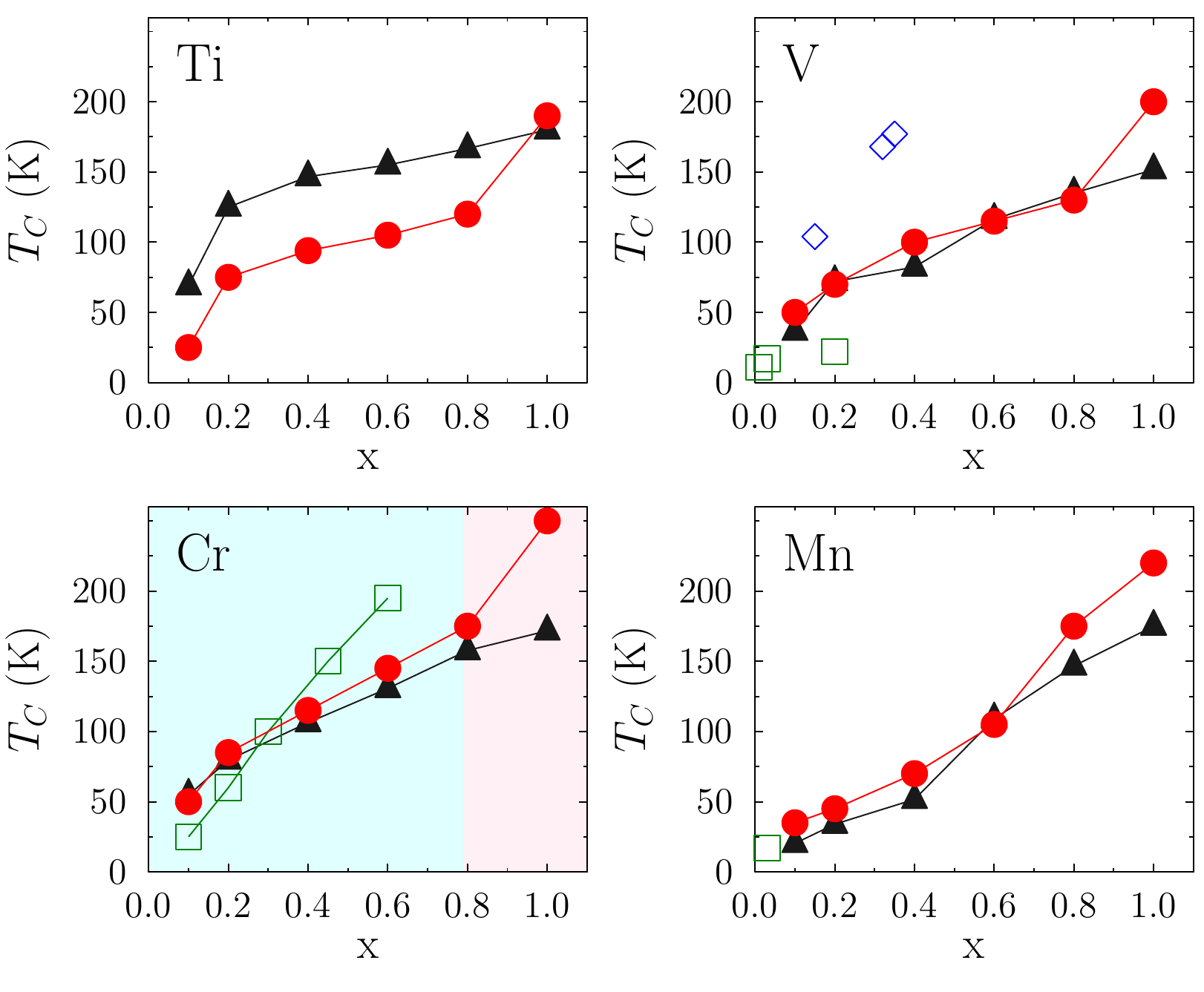}
\caption{(color online) Calculated critical temperature
  $T_{\mathrm{C}}$ versus concentration $x$ of Ti, V, Cr, and Mn in
  Sb$_{2-x}$TM$_{x}$Te$_3$. The critical temperature $T_{\mathrm{C}}$
  is determined from Monte Carlo simulations with randomly distributed
  impurities (black filled triangles) and with cluster percolation
  (red filled circles). $T_{\mathrm{C}}$ is compared to experimental
  data (green and blue markers) \cite{Zhou2005,Zhou2006a}. In the case
  of Cr, there is a transition from a ferromagnetic state for $x<0.8$
  (light blue background) to an antiferromagnetic state at higher
  concentrations (light red background).}
\label{fig:4}
\end{figure}

Considering a wider range of concentrations of the TM atom, we have estimated the critical temperatures $T_{\mathrm{C}}$ using a Monte
Carlo method \cite{Metropolis1953,Binder1997,Fischer2009}. To treat both
ferromagnetic and antiferromagnetic materials, we investigate the
spin-spin-correlation function
\begin{equation*}
 S=\sum_{i}\sum_{j\in\Omega_{i}}\left|\vec{m}_{i}\cdot\vec{m}_{j}\right|,
\end{equation*}
where $\vec{m}_{i}$ and $\Omega_{i}$ are the magnetic moment and the
interaction sphere around site $i$, respectively.  We also account for
percolation effects, using pair potentials, and compared estimated
critical temperatures with the available experimental data.  The
results for ferromagnetic Sb$_{2-x}$TM$_{x}$Te$_3$ (TM = Ti, V, Cr,
Mn; Fig.~\ref{fig:4}) show a systematic increase of the $T_{\mathrm{C}}$ with the concentration of dopants. Percolation
effects do not affect strongly the behavior of $T_{\mathrm{C}}$ at low
concentrations; except in the case of Ti, for which percolation lowers
$T_{\mathrm{C}}$. Calculations for Cr reproduce the experimentally measured trends
for concentrations up to $x=0.6$
\cite{Zhou2006a}. For higher concentrations, we found a transition to
antiferromagnetic order (area with a light red background in
Fig.~\ref{fig:4}), which is understood as the results of an increasing
antiferromagnetic interaction between magnetic moments from nearby quintuple layers. This explains why experimental data is
unavailable for concentrations larger than $x=0.6$.

Concerning the Sb$_{2-x}$V$_{x}$Te$_3$ case, we reproduce the trend
found in experiments for a broad concentration range \cite{Zhou2005}.
However, the theoretical absolute values are underestimated by about a
factor of $1/2$. One could speculate that either structural imperfections due
to sample preparations or limitations in the \textit{ab initio}
description could cause this mismatch. For the
reported cases of Mn doping at low concentration regimes ($x\le 0.1$),
our calculations are, again, in qualitative agreement with experiment
\cite{Dyck2003,Choi2004,Choi2005,Choi2006,Choi2012,Hor2010}.

The systematic study of exchange interaction discussed for TM-doped magnetic chalcogenides
elucidated in this Letter can open new possibilities for a specific design
of their magnetic properties. We infer for instance that one way to control the magnetic
interaction and the Curie temperature is to replace particular atoms
or sheets of atoms, in order to tune the strength of the electronic
hybridization; the latter is responsible for the exchange interaction
mechanisms in this class of materials. Here, it has been shown that
the overlap between electronic wave functions of anions and cations is
crucial.

In an experimental realization, one could replace the anion layer
between two cation sheets by atoms of the same group in the periodic
table.  As an example, we consider the layer between two
Sb$_{1.80}$Cr$_{0.20}$ sheets in the Sb$_{1.80}$Cr$_{0.20}$Te$_3$
alternated with various chalcogen atoms. We observe an increase of the associated Curie
temperature with ionic size and spatial extension of the
electronic wave function of the chalcogen ion: S, Se, Te, and Po yield
$T_{\mathrm{C}} = 69$ K, $71$ K, $76$ K, and $79$ K, respectively. A
similar effect can be achieved with specific co-doping of the
corresponding anion layer.

The in-plane exchange interaction can instead be tuned by co-doping of
the cation layers in accordance with the exchange constant behavior
presented in Fig.~\ref{fig:2}. To illustrate this aspect, we calculated the
exchange interaction in Bi$_{2-x-y}$Cr$_{x}$Sb$_{y}$Te$_3$ for
$x=0.2$ and $y=0.0$, $0.2$, and $0.4$ (detailed results are presented in the
Supplementary Material). Here, the Curie temperature increases with Sb
concentration from $35$ K at $y = 0.0$ to $39$ K at $y = 0.2$, and to
$42$ K at $y = 0.4$. 

Another way to tune the magnetic interaction is to insert impurities
into the van der Waals gap \cite{Otrokov2013}. This can change the
size of the van der Waals gap and, by a proper choice of the
impurities, can supply free carriers, which are important for the
indirect exchange of Zener type (see  the
Supplementary Material).

In summary, we have studied the exchange interaction in the
Bi$_2$Te$_3$, Bi$_2$Se$_3$, and Sb$_2$Te$_3$ tetradymite chalcogenides
doped with transition metals. Our first principles
calculations have shown that the magnetic interaction is long-range
and is mainly mediated out-of-plane by the double exchange mechanism
via an anion and in-plane by the indirect exchange coupling via free
carriers. The calculated Curie temperatures as a function of the
concentration are found in qualitative agreement with available
experimental data.  Finally, we presented several ways to tune the
magnetic interaction in these systems: (i) replacing the anion layer
between two cation sheets by atoms of the same group, (ii) co-doping
of the cation sheet, and (iii) inserting impurities in the van der
Waals gap.  These results provide deep insight into the magnetic
interactions in the magnetic binary chalcogenides, and open new ways to
design new materials for promising applications.

We acknowledge support by the Ministry of Education and Science of
Russian Federation (state task No. 2.8575.2013) and the Deutsche
Forschungsgemeinschaft (Priority Program SPP 1666 ''Topological
Insulators''). The calculations were performed at the Rechenzentrum
Garching of the Max Planck Society (Germany) and at the SKIF-Cyberia
supercomputer of Tomsk State University.
\bibliographystyle{apsrev}
\bibliography{./chalcogenides}

\begin{thebibliography}{43}
\expandafter\ifx\csname natexlab\endcsname\relax\def\natexlab#1{#1}\fi
\expandafter\ifx\csname bibnamefont\endcsname\relax
  \def\bibnamefont#1{#1}\fi
\expandafter\ifx\csname bibfnamefont\endcsname\relax
  \def\bibfnamefont#1{#1}\fi
\expandafter\ifx\csname citenamefont\endcsname\relax
  \def\citenamefont#1{#1}\fi
\expandafter\ifx\csname url\endcsname\relax
  \def\url#1{\texttt{#1}}\fi
\expandafter\ifx\csname urlprefix\endcsname\relax\def\urlprefix{URL }\fi
\providecommand{\bibinfo}[2]{#2}
\providecommand{\eprint}[2][]{\url{#2}}

\bibitem[{\citenamefont{Nolas et~al.}(2001)\citenamefont{Nolas, Sharp, and
  Goldsmid}}]{Nolas2001}
\bibinfo{author}{\bibfnamefont{G.}~\bibnamefont{Nolas}},
  \bibinfo{author}{\bibfnamefont{J.}~\bibnamefont{Sharp}}, \bibnamefont{and}
  \bibinfo{author}{\bibfnamefont{J.}~\bibnamefont{Goldsmid}},
  \emph{\bibinfo{title}{Thermoelectrics: Basic Principles and New Materials
  Developments}}, Springer Series in Materials Science
  (\bibinfo{publisher}{Springer}, \bibinfo{year}{2001}).

\bibitem[{\citenamefont{Hasan and Kane}(2010)}]{Hasan2010}
\bibinfo{author}{\bibfnamefont{M.~Z.} \bibnamefont{Hasan}} \bibnamefont{and}
  \bibinfo{author}{\bibfnamefont{C.~L.} \bibnamefont{Kane}},
  \bibinfo{journal}{Rev. Mod. Phys.} \textbf{\bibinfo{volume}{82}},
  \bibinfo{pages}{3045} (\bibinfo{year}{2010}).

\bibitem[{\citenamefont{Qi and Zhang}(2011)}]{Qi2011}
\bibinfo{author}{\bibfnamefont{X.-L.} \bibnamefont{Qi}} \bibnamefont{and}
  \bibinfo{author}{\bibfnamefont{S.-C.} \bibnamefont{Zhang}},
  \bibinfo{journal}{Rev. Mod. Phys.} \textbf{\bibinfo{volume}{83}},
  \bibinfo{pages}{1057} (\bibinfo{year}{2011}).

\bibitem[{\citenamefont{Eremeev et~al.}(2012)\citenamefont{Eremeev, Landolt,
  Menshchikova, Slomski, Koroteev, Aliev, Babanly, Henk, Ernst, Patthey
  et~al.}}]{Eremeev2012}
\bibinfo{author}{\bibfnamefont{S.~V.} \bibnamefont{Eremeev}},
  \bibinfo{author}{\bibfnamefont{G.}~\bibnamefont{Landolt}},
  \bibinfo{author}{\bibfnamefont{T.~V.} \bibnamefont{Menshchikova}},
  \bibinfo{author}{\bibfnamefont{B.}~\bibnamefont{Slomski}},
  \bibinfo{author}{\bibfnamefont{Y.~M.} \bibnamefont{Koroteev}},
  \bibinfo{author}{\bibfnamefont{Z.~S.} \bibnamefont{Aliev}},
  \bibinfo{author}{\bibfnamefont{M.~B.} \bibnamefont{Babanly}},
  \bibinfo{author}{\bibfnamefont{J.}~\bibnamefont{Henk}},
  \bibinfo{author}{\bibfnamefont{A.}~\bibnamefont{Ernst}},
  \bibinfo{author}{\bibfnamefont{L.}~\bibnamefont{Patthey}},
  \bibnamefont{et~al.}, \bibinfo{journal}{Nat Commun}
  \textbf{\bibinfo{volume}{3}}, \bibinfo{pages}{635} (\bibinfo{year}{2012}).

\bibitem[{\citenamefont{\v{Z}uti\'{c} et~al.}(2004)\citenamefont{\v{Z}uti\'{c},
  Fabian, and Das~Sarma}}]{Zutic2004}
\bibinfo{author}{\bibfnamefont{I.}~\bibnamefont{\v{Z}uti\'{c}}},
  \bibinfo{author}{\bibfnamefont{J.}~\bibnamefont{Fabian}}, \bibnamefont{and}
  \bibinfo{author}{\bibfnamefont{S.}~\bibnamefont{Das~Sarma}},
  \bibinfo{journal}{Rev. Mod. Phys.} \textbf{\bibinfo{volume}{76}},
  \bibinfo{pages}{323} (\bibinfo{year}{2004}).

\bibitem[{\citenamefont{Sato et~al.}(2010)\citenamefont{Sato, Bergqvist,
  Kudrnovsk\'y, Dederichs, Eriksson, Turek, Sanyal, Bouzerar, Katayama-Yoshida,
  Dinh et~al.}}]{Sato2010}
\bibinfo{author}{\bibfnamefont{K.}~\bibnamefont{Sato}},
  \bibinfo{author}{\bibfnamefont{L.}~\bibnamefont{Bergqvist}},
  \bibinfo{author}{\bibfnamefont{J.}~\bibnamefont{Kudrnovsk\'y}},
  \bibinfo{author}{\bibfnamefont{P.~H.} \bibnamefont{Dederichs}},
  \bibinfo{author}{\bibfnamefont{O.}~\bibnamefont{Eriksson}},
  \bibinfo{author}{\bibfnamefont{I.}~\bibnamefont{Turek}},
  \bibinfo{author}{\bibfnamefont{B.}~\bibnamefont{Sanyal}},
  \bibinfo{author}{\bibfnamefont{G.}~\bibnamefont{Bouzerar}},
  \bibinfo{author}{\bibfnamefont{H.}~\bibnamefont{Katayama-Yoshida}},
  \bibinfo{author}{\bibfnamefont{V.~A.} \bibnamefont{Dinh}},
  \bibnamefont{et~al.}, \bibinfo{journal}{Rev. Mod. Phys.}
  \textbf{\bibinfo{volume}{82}}, \bibinfo{pages}{1633} (\bibinfo{year}{2010}).

\bibitem[{\citenamefont{Uchida et~al.}(2008)\citenamefont{Uchida, Takahashi,
  Harii, Ieda, Koshibae, Ando, Maekawa, and Saitoh}}]{Uchida2008}
\bibinfo{author}{\bibfnamefont{K.}~\bibnamefont{Uchida}},
  \bibinfo{author}{\bibfnamefont{S.}~\bibnamefont{Takahashi}},
  \bibinfo{author}{\bibfnamefont{K.}~\bibnamefont{Harii}},
  \bibinfo{author}{\bibfnamefont{J.}~\bibnamefont{Ieda}},
  \bibinfo{author}{\bibfnamefont{W.}~\bibnamefont{Koshibae}},
  \bibinfo{author}{\bibfnamefont{K.}~\bibnamefont{Ando}},
  \bibinfo{author}{\bibfnamefont{S.}~\bibnamefont{Maekawa}}, \bibnamefont{and}
  \bibinfo{author}{\bibfnamefont{E.}~\bibnamefont{Saitoh}},
  \bibinfo{journal}{Nature} \textbf{\bibinfo{volume}{455}},
  \bibinfo{pages}{778} (\bibinfo{year}{2008}).

\bibitem[{\citenamefont{Bauer et~al.}(2012)\citenamefont{Bauer, Saitoh, and van
  Wees}}]{Bauer2012}
\bibinfo{author}{\bibfnamefont{G.~E.~W.} \bibnamefont{Bauer}},
  \bibinfo{author}{\bibfnamefont{E.}~\bibnamefont{Saitoh}}, \bibnamefont{and}
  \bibinfo{author}{\bibfnamefont{B.~J.} \bibnamefont{van Wees}},
  \bibinfo{journal}{Nat Mater} \textbf{\bibinfo{volume}{11}},
  \bibinfo{pages}{391} (\bibinfo{year}{2012}).

\bibitem[{\citenamefont{Dyck et~al.}(2002)\citenamefont{Dyck, Hajek, Lostak,
  and Uher}}]{Dyck2002}
\bibinfo{author}{\bibfnamefont{J.~S.} \bibnamefont{Dyck}},
  \bibinfo{author}{\bibfnamefont{P.}~\bibnamefont{Hajek}},
  \bibinfo{author}{\bibfnamefont{P.}~\bibnamefont{Lostak}}, \bibnamefont{and}
  \bibinfo{author}{\bibfnamefont{C.}~\bibnamefont{Uher}},
  \bibinfo{journal}{Phys. Rev. B} \textbf{\bibinfo{volume}{65}},
  \bibinfo{pages}{115212} (\bibinfo{year}{2002}).

\bibitem[{\citenamefont{Dyck et~al.}(2003)\citenamefont{Dyck, Svanda, Lostak,
  Horak, Chen, and Uher}}]{Dyck2003}
\bibinfo{author}{\bibfnamefont{J.~S.} \bibnamefont{Dyck}},
  \bibinfo{author}{\bibfnamefont{P.}~\bibnamefont{Svanda}},
  \bibinfo{author}{\bibfnamefont{P.}~\bibnamefont{Lostak}},
  \bibinfo{author}{\bibfnamefont{J.}~\bibnamefont{Horak}},
  \bibinfo{author}{\bibfnamefont{W.}~\bibnamefont{Chen}}, \bibnamefont{and}
  \bibinfo{author}{\bibfnamefont{C.}~\bibnamefont{Uher}},
  \bibinfo{journal}{Journal of Applied Physics} \textbf{\bibinfo{volume}{94}},
  \bibinfo{pages}{7631} (\bibinfo{year}{2003}).

\bibitem[{\citenamefont{Dyck et~al.}(2005)\citenamefont{Dyck, Drasar, Lostak,
  and Uher}}]{Dyck2005}
\bibinfo{author}{\bibfnamefont{J.~S.} \bibnamefont{Dyck}},
  \bibinfo{author}{\bibfnamefont{C.}~\bibnamefont{Drasar}},
  \bibinfo{author}{\bibfnamefont{P.}~\bibnamefont{Lostak}}, \bibnamefont{and}
  \bibinfo{author}{\bibfnamefont{C.}~\bibnamefont{Uher}},
  \bibinfo{journal}{Phys. Rev. B} \textbf{\bibinfo{volume}{71}},
  \bibinfo{pages}{115214} (\bibinfo{year}{2005}).

\bibitem[{\citenamefont{Kulbachinskii et~al.}(2003)\citenamefont{Kulbachinskii,
  Kaminsky, Kindo, Narumi, Suga, Lostak, and Svanda}}]{Kulbachinskii2003}
\bibinfo{author}{\bibfnamefont{V.~A.} \bibnamefont{Kulbachinskii}},
  \bibinfo{author}{\bibfnamefont{A.~Y.} \bibnamefont{Kaminsky}},
  \bibinfo{author}{\bibfnamefont{K.}~\bibnamefont{Kindo}},
  \bibinfo{author}{\bibfnamefont{Y.}~\bibnamefont{Narumi}},
  \bibinfo{author}{\bibfnamefont{K.-i.} \bibnamefont{Suga}},
  \bibinfo{author}{\bibfnamefont{P.}~\bibnamefont{Lostak}}, \bibnamefont{and}
  \bibinfo{author}{\bibfnamefont{P.}~\bibnamefont{Svanda}},
  \bibinfo{journal}{Physica B: Condensed Matter}
  \textbf{\bibinfo{volume}{329-333, Part 2}}, \bibinfo{pages}{1251}
  (\bibinfo{year}{2003}).

\bibitem[{\citenamefont{Kulbachinskii et~al.}(2005)\citenamefont{Kulbachinskii,
  Tarasov, and Brück}}]{Kulbachinskii2005}
\bibinfo{author}{\bibfnamefont{V.}~\bibnamefont{Kulbachinskii}},
  \bibinfo{author}{\bibfnamefont{P.}~\bibnamefont{Tarasov}}, \bibnamefont{and}
  \bibinfo{author}{\bibfnamefont{E.}~\bibnamefont{Brück}},
  \bibinfo{journal}{Journal of Experimental and Theoretical Physics}
  \textbf{\bibinfo{volume}{101}}, \bibinfo{pages}{528} (\bibinfo{year}{2005}).

\bibitem[{\citenamefont{Choi et~al.}(2004)\citenamefont{Choi, Choi, Choi, Park,
  Park, Lee, Woo, and Cho}}]{Choi2004}
\bibinfo{author}{\bibfnamefont{J.}~\bibnamefont{Choi}},
  \bibinfo{author}{\bibfnamefont{S.}~\bibnamefont{Choi}},
  \bibinfo{author}{\bibfnamefont{J.}~\bibnamefont{Choi}},
  \bibinfo{author}{\bibfnamefont{Y.}~\bibnamefont{Park}},
  \bibinfo{author}{\bibfnamefont{H.-M.} \bibnamefont{Park}},
  \bibinfo{author}{\bibfnamefont{H.-W.} \bibnamefont{Lee}},
  \bibinfo{author}{\bibfnamefont{B.-C.} \bibnamefont{Woo}}, \bibnamefont{and}
  \bibinfo{author}{\bibfnamefont{S.}~\bibnamefont{Cho}},
  \bibinfo{journal}{Phys. Stat. Sol. (b)} \textbf{\bibinfo{volume}{241}},
  \bibinfo{pages}{1541} (\bibinfo{year}{2004}).

\bibitem[{\citenamefont{Zhou et~al.}(2005)\citenamefont{Zhou, Chien, and
  Uher}}]{Zhou2005}
\bibinfo{author}{\bibfnamefont{Z.}~\bibnamefont{Zhou}},
  \bibinfo{author}{\bibfnamefont{Y.-J.} \bibnamefont{Chien}}, \bibnamefont{and}
  \bibinfo{author}{\bibfnamefont{C.}~\bibnamefont{Uher}},
  \bibinfo{journal}{Applied Physics Letters} \textbf{\bibinfo{volume}{87}},
  \bibinfo{pages}{112503} (\bibinfo{year}{2005}).

\bibitem[{\citenamefont{Zhou et~al.}(2006{\natexlab{a}})\citenamefont{Zhou,
  Chien, and Uher}}]{Zhou2006a}
\bibinfo{author}{\bibfnamefont{Z.}~\bibnamefont{Zhou}},
  \bibinfo{author}{\bibfnamefont{Y.-J.} \bibnamefont{Chien}}, \bibnamefont{and}
  \bibinfo{author}{\bibfnamefont{C.}~\bibnamefont{Uher}},
  \bibinfo{journal}{Phys. Rev. B} \textbf{\bibinfo{volume}{74}},
  \bibinfo{pages}{224418} (\bibinfo{year}{2006}{\natexlab{a}}).

\bibitem[{\citenamefont{Zhou et~al.}(2006{\natexlab{b}})\citenamefont{Zhou,
  Zabeik, Lostak, and Uher}}]{Zhou2006c}
\bibinfo{author}{\bibfnamefont{Z.}~\bibnamefont{Zhou}},
  \bibinfo{author}{\bibfnamefont{M.}~\bibnamefont{Zabeik}},
  \bibinfo{author}{\bibfnamefont{P.}~\bibnamefont{Lostak}}, \bibnamefont{and}
  \bibinfo{author}{\bibfnamefont{C.}~\bibnamefont{Uher}},
  \bibinfo{journal}{Journal of Applied Physics} \textbf{\bibinfo{volume}{99}},
  \bibinfo{pages}{043901} (\bibinfo{year}{2006}{\natexlab{b}}).

\bibitem[{\citenamefont{Bos et~al.}(2006)\citenamefont{Bos, Lee, Morosan,
  Zandbergen, Lee, Ong, and Cava}}]{Bos2006}
\bibinfo{author}{\bibfnamefont{J.~W.~G.} \bibnamefont{Bos}},
  \bibinfo{author}{\bibfnamefont{M.}~\bibnamefont{Lee}},
  \bibinfo{author}{\bibfnamefont{E.}~\bibnamefont{Morosan}},
  \bibinfo{author}{\bibfnamefont{H.~W.} \bibnamefont{Zandbergen}},
  \bibinfo{author}{\bibfnamefont{W.~L.} \bibnamefont{Lee}},
  \bibinfo{author}{\bibfnamefont{N.~P.} \bibnamefont{Ong}}, \bibnamefont{and}
  \bibinfo{author}{\bibfnamefont{R.~J.} \bibnamefont{Cava}},
  \bibinfo{journal}{Phys. Rev. B} \textbf{\bibinfo{volume}{74}},
  \bibinfo{pages}{184429} (\bibinfo{year}{2006}).

\bibitem[{\citenamefont{Hor et~al.}(2010)\citenamefont{Hor, Roushan,
  Beidenkopf, Seo, Qu, Checkelsky, Wray, Hsieh, Xia, Xu et~al.}}]{Hor2010}
\bibinfo{author}{\bibfnamefont{Y.~S.} \bibnamefont{Hor}},
  \bibinfo{author}{\bibfnamefont{P.}~\bibnamefont{Roushan}},
  \bibinfo{author}{\bibfnamefont{H.}~\bibnamefont{Beidenkopf}},
  \bibinfo{author}{\bibfnamefont{J.}~\bibnamefont{Seo}},
  \bibinfo{author}{\bibfnamefont{D.}~\bibnamefont{Qu}},
  \bibinfo{author}{\bibfnamefont{J.~G.} \bibnamefont{Checkelsky}},
  \bibinfo{author}{\bibfnamefont{L.~A.} \bibnamefont{Wray}},
  \bibinfo{author}{\bibfnamefont{D.}~\bibnamefont{Hsieh}},
  \bibinfo{author}{\bibfnamefont{Y.}~\bibnamefont{Xia}},
  \bibinfo{author}{\bibfnamefont{S.-Y.} \bibnamefont{Xu}},
  \bibnamefont{et~al.}, \bibinfo{journal}{Phys. Rev. B}
  \textbf{\bibinfo{volume}{81}}, \bibinfo{pages}{195203}
  (\bibinfo{year}{2010}).

\bibitem[{\citenamefont{Hsieh et~al.}(2009)\citenamefont{Hsieh, Xia, Qian,
  Wray, Meier, Dil, Osterwalder, Patthey, Fedorov, Lin et~al.}}]{Hsieh2009b}
\bibinfo{author}{\bibfnamefont{D.}~\bibnamefont{Hsieh}},
  \bibinfo{author}{\bibfnamefont{Y.}~\bibnamefont{Xia}},
  \bibinfo{author}{\bibfnamefont{D.}~\bibnamefont{Qian}},
  \bibinfo{author}{\bibfnamefont{L.}~\bibnamefont{Wray}},
  \bibinfo{author}{\bibfnamefont{F.}~\bibnamefont{Meier}},
  \bibinfo{author}{\bibfnamefont{J.~H.} \bibnamefont{Dil}},
  \bibinfo{author}{\bibfnamefont{J.}~\bibnamefont{Osterwalder}},
  \bibinfo{author}{\bibfnamefont{L.}~\bibnamefont{Patthey}},
  \bibinfo{author}{\bibfnamefont{A.~V.} \bibnamefont{Fedorov}},
  \bibinfo{author}{\bibfnamefont{H.}~\bibnamefont{Lin}}, \bibnamefont{et~al.},
  \bibinfo{journal}{Phys. Rev. Lett.} \textbf{\bibinfo{volume}{103}},
  \bibinfo{pages}{146401} (\bibinfo{year}{2009}).

\bibitem[{\citenamefont{Honolka et~al.}(2012)\citenamefont{Honolka,
  Khajetoorians, Sessi, Wehling, Stepanow, Mi, Iversen, Schlenk, Wiebe, Brookes
  et~al.}}]{Honolka2012}
\bibinfo{author}{\bibfnamefont{J.}~\bibnamefont{Honolka}},
  \bibinfo{author}{\bibfnamefont{A.~A.} \bibnamefont{Khajetoorians}},
  \bibinfo{author}{\bibfnamefont{V.}~\bibnamefont{Sessi}},
  \bibinfo{author}{\bibfnamefont{T.~O.} \bibnamefont{Wehling}},
  \bibinfo{author}{\bibfnamefont{S.}~\bibnamefont{Stepanow}},
  \bibinfo{author}{\bibfnamefont{J.-L.} \bibnamefont{Mi}},
  \bibinfo{author}{\bibfnamefont{B.~B.} \bibnamefont{Iversen}},
  \bibinfo{author}{\bibfnamefont{T.}~\bibnamefont{Schlenk}},
  \bibinfo{author}{\bibfnamefont{J.}~\bibnamefont{Wiebe}},
  \bibinfo{author}{\bibfnamefont{N.~B.} \bibnamefont{Brookes}},
  \bibnamefont{et~al.}, \bibinfo{journal}{Phys. Rev. Lett.}
  \textbf{\bibinfo{volume}{108}}, \bibinfo{pages}{256811}
  (\bibinfo{year}{2012}).

\bibitem[{\citenamefont{Shelford et~al.}(2012)\citenamefont{Shelford, Hesjedal,
  Collins-McIntyre, Dhesi, Maccherozzi, and van~der Laan}}]{Shelford2012}
\bibinfo{author}{\bibfnamefont{L.~R.} \bibnamefont{Shelford}},
  \bibinfo{author}{\bibfnamefont{T.}~\bibnamefont{Hesjedal}},
  \bibinfo{author}{\bibfnamefont{L.}~\bibnamefont{Collins-McIntyre}},
  \bibinfo{author}{\bibfnamefont{S.~S.} \bibnamefont{Dhesi}},
  \bibinfo{author}{\bibfnamefont{F.}~\bibnamefont{Maccherozzi}},
  \bibnamefont{and} \bibinfo{author}{\bibfnamefont{G.}~\bibnamefont{van~der
  Laan}}, \bibinfo{journal}{Phys. Rev. B} \textbf{\bibinfo{volume}{86}},
  \bibinfo{pages}{081304} (\bibinfo{year}{2012}).

\bibitem[{\citenamefont{West et~al.}(2012)\citenamefont{West, Sun, Zhang,
  Zhang, Ma, Cheng, Zhang, Chen, Jia, and Xue}}]{West2012}
\bibinfo{author}{\bibfnamefont{D.}~\bibnamefont{West}},
  \bibinfo{author}{\bibfnamefont{Y.~Y.} \bibnamefont{Sun}},
  \bibinfo{author}{\bibfnamefont{S.~B.} \bibnamefont{Zhang}},
  \bibinfo{author}{\bibfnamefont{T.}~\bibnamefont{Zhang}},
  \bibinfo{author}{\bibfnamefont{X.}~\bibnamefont{Ma}},
  \bibinfo{author}{\bibfnamefont{P.}~\bibnamefont{Cheng}},
  \bibinfo{author}{\bibfnamefont{Y.~Y.} \bibnamefont{Zhang}},
  \bibinfo{author}{\bibfnamefont{X.}~\bibnamefont{Chen}},
  \bibinfo{author}{\bibfnamefont{J.~F.} \bibnamefont{Jia}}, \bibnamefont{and}
  \bibinfo{author}{\bibfnamefont{Q.~K.} \bibnamefont{Xue}},
  \bibinfo{journal}{Phys. Rev. B} \textbf{\bibinfo{volume}{85}},
  \bibinfo{pages}{081305} (\bibinfo{year}{2012}).

\bibitem[{\citenamefont{Scholz et~al.}(2012)\citenamefont{Scholz,
  S\'anchez-Barriga, Marchenko, Varykhalov, Volykhov, Yashina, and
  Rader}}]{Scholz2012}
\bibinfo{author}{\bibfnamefont{M.~R.} \bibnamefont{Scholz}},
  \bibinfo{author}{\bibfnamefont{J.}~\bibnamefont{S\'anchez-Barriga}},
  \bibinfo{author}{\bibfnamefont{D.}~\bibnamefont{Marchenko}},
  \bibinfo{author}{\bibfnamefont{A.}~\bibnamefont{Varykhalov}},
  \bibinfo{author}{\bibfnamefont{A.}~\bibnamefont{Volykhov}},
  \bibinfo{author}{\bibfnamefont{L.~V.} \bibnamefont{Yashina}},
  \bibnamefont{and} \bibinfo{author}{\bibfnamefont{O.}~\bibnamefont{Rader}},
  \bibinfo{journal}{Phys. Rev. Lett.} \textbf{\bibinfo{volume}{108}},
  \bibinfo{pages}{256810} (\bibinfo{year}{2012}).

\bibitem[{\citenamefont{Kou et~al.}(2012)\citenamefont{Kou, Jiang, Lang, Xiu,
  He, Wang, Wang, Yu, Fedorov, Zhang et~al.}}]{Kou2012}
\bibinfo{author}{\bibfnamefont{X.~F.} \bibnamefont{Kou}},
  \bibinfo{author}{\bibfnamefont{W.~J.} \bibnamefont{Jiang}},
  \bibinfo{author}{\bibfnamefont{M.~R.} \bibnamefont{Lang}},
  \bibinfo{author}{\bibfnamefont{F.~X.} \bibnamefont{Xiu}},
  \bibinfo{author}{\bibfnamefont{L.}~\bibnamefont{He}},
  \bibinfo{author}{\bibfnamefont{Y.}~\bibnamefont{Wang}},
  \bibinfo{author}{\bibfnamefont{Y.}~\bibnamefont{Wang}},
  \bibinfo{author}{\bibfnamefont{X.~X.} \bibnamefont{Yu}},
  \bibinfo{author}{\bibfnamefont{A.~V.} \bibnamefont{Fedorov}},
  \bibinfo{author}{\bibfnamefont{P.}~\bibnamefont{Zhang}},
  \bibnamefont{et~al.}, \bibinfo{journal}{Journal of Applied Physics}
  \textbf{\bibinfo{volume}{112}}, \bibinfo{pages}{063912}
  (\bibinfo{year}{2012}).

\bibitem[{\citenamefont{Ye et~al.}(2012)\citenamefont{Ye, Eremeev, Kuroda,
  Krasovskii, Chulkov, Takeda, Saitoh, Okamoto, Zhu, Miyamoto et~al.}}]{Ye2012}
\bibinfo{author}{\bibfnamefont{M.}~\bibnamefont{Ye}},
  \bibinfo{author}{\bibfnamefont{S.~V.} \bibnamefont{Eremeev}},
  \bibinfo{author}{\bibfnamefont{K.}~\bibnamefont{Kuroda}},
  \bibinfo{author}{\bibfnamefont{E.~E.} \bibnamefont{Krasovskii}},
  \bibinfo{author}{\bibfnamefont{E.~V.} \bibnamefont{Chulkov}},
  \bibinfo{author}{\bibfnamefont{Y.}~\bibnamefont{Takeda}},
  \bibinfo{author}{\bibfnamefont{Y.}~\bibnamefont{Saitoh}},
  \bibinfo{author}{\bibfnamefont{K.}~\bibnamefont{Okamoto}},
  \bibinfo{author}{\bibfnamefont{S.~Y.} \bibnamefont{Zhu}},
  \bibinfo{author}{\bibfnamefont{K.}~\bibnamefont{Miyamoto}},
  \bibnamefont{et~al.}, \bibinfo{journal}{Phys. Rev. B}
  \textbf{\bibinfo{volume}{85}}, \bibinfo{pages}{205317}
  (\bibinfo{year}{2012}).

\bibitem[{\citenamefont{Larson and Lambrecht}(2008)}]{Larson2008}
\bibinfo{author}{\bibfnamefont{P.}~\bibnamefont{Larson}} \bibnamefont{and}
  \bibinfo{author}{\bibfnamefont{W.~R.~L.} \bibnamefont{Lambrecht}},
  \bibinfo{journal}{Phys. Rev. B} \textbf{\bibinfo{volume}{78}},
  \bibinfo{pages}{195207} (\bibinfo{year}{2008}).

\bibitem[{\citenamefont{Yu et~al.}(2010)\citenamefont{Yu, Zhang, Zhang, Zhang,
  Dai, and Fang}}]{Yu2010}
\bibinfo{author}{\bibfnamefont{R.}~\bibnamefont{Yu}},
  \bibinfo{author}{\bibfnamefont{W.}~\bibnamefont{Zhang}},
  \bibinfo{author}{\bibfnamefont{H.-J.} \bibnamefont{Zhang}},
  \bibinfo{author}{\bibfnamefont{S.-C.} \bibnamefont{Zhang}},
  \bibinfo{author}{\bibfnamefont{X.}~\bibnamefont{Dai}}, \bibnamefont{and}
  \bibinfo{author}{\bibfnamefont{Z.}~\bibnamefont{Fang}},
  \bibinfo{journal}{Science} \textbf{\bibinfo{volume}{329}},
  \bibinfo{pages}{61} (\bibinfo{year}{2010}).

\bibitem[{\citenamefont{Zhang et~al.}(2012)\citenamefont{Zhang, Zhu, Zhang,
  Xiao, and Yao}}]{Zhang2012}
\bibinfo{author}{\bibfnamefont{J.-M.} \bibnamefont{Zhang}},
  \bibinfo{author}{\bibfnamefont{W.}~\bibnamefont{Zhu}},
  \bibinfo{author}{\bibfnamefont{Y.}~\bibnamefont{Zhang}},
  \bibinfo{author}{\bibfnamefont{D.}~\bibnamefont{Xiao}}, \bibnamefont{and}
  \bibinfo{author}{\bibfnamefont{Y.}~\bibnamefont{Yao}},
  \bibinfo{journal}{Phys. Rev. Lett.} \textbf{\bibinfo{volume}{109}},
  \bibinfo{pages}{266405} (\bibinfo{year}{2012}).

\bibitem[{\citenamefont{Henk et~al.}(2012{\natexlab{a}})\citenamefont{Henk,
  Ernst, Eremeev, Chulkov, Maznichenko, and Mertig}}]{Henk2012}
\bibinfo{author}{\bibfnamefont{J.}~\bibnamefont{Henk}},
  \bibinfo{author}{\bibfnamefont{A.}~\bibnamefont{Ernst}},
  \bibinfo{author}{\bibfnamefont{S.~V.} \bibnamefont{Eremeev}},
  \bibinfo{author}{\bibfnamefont{E.~V.} \bibnamefont{Chulkov}},
  \bibinfo{author}{\bibfnamefont{I.~V.} \bibnamefont{Maznichenko}},
  \bibnamefont{and} \bibinfo{author}{\bibfnamefont{I.}~\bibnamefont{Mertig}},
  \bibinfo{journal}{Phys. Rev. Lett.} \textbf{\bibinfo{volume}{108}},
  \bibinfo{pages}{206801} (\bibinfo{year}{2012}{\natexlab{a}}).

\bibitem[{\citenamefont{Henk et~al.}(2012{\natexlab{b}})\citenamefont{Henk,
  Flieger, Maznichenko, Mertig, Ernst, Eremeev, and Chulkov}}]{Henk2012a}
\bibinfo{author}{\bibfnamefont{J.}~\bibnamefont{Henk}},
  \bibinfo{author}{\bibfnamefont{M.}~\bibnamefont{Flieger}},
  \bibinfo{author}{\bibfnamefont{I.~V.} \bibnamefont{Maznichenko}},
  \bibinfo{author}{\bibfnamefont{I.}~\bibnamefont{Mertig}},
  \bibinfo{author}{\bibfnamefont{A.}~\bibnamefont{Ernst}},
  \bibinfo{author}{\bibfnamefont{S.~V.} \bibnamefont{Eremeev}},
  \bibnamefont{and} \bibinfo{author}{\bibfnamefont{E.~V.}
  \bibnamefont{Chulkov}}, \bibinfo{journal}{Phys. Rev. Lett.}
  \textbf{\bibinfo{volume}{109}}, \bibinfo{pages}{076801}
  (\bibinfo{year}{2012}{\natexlab{b}}).

\bibitem[{\citenamefont{Perdew and Wang}(1992)}]{Perdew1992a}
\bibinfo{author}{\bibfnamefont{J.~P.} \bibnamefont{Perdew}} \bibnamefont{and}
  \bibinfo{author}{\bibfnamefont{Y.}~\bibnamefont{Wang}},
  \bibinfo{journal}{Phys. Rev. B} \textbf{\bibinfo{volume}{45}},
  \bibinfo{pages}{13244} (\bibinfo{year}{1992}).

\bibitem[{\citenamefont{Perdew et~al.}(1996)\citenamefont{Perdew, Burke, and
  Ernzerhof}}]{Perdew1996}
\bibinfo{author}{\bibfnamefont{J.~P.} \bibnamefont{Perdew}},
  \bibinfo{author}{\bibfnamefont{K.}~\bibnamefont{Burke}}, \bibnamefont{and}
  \bibinfo{author}{\bibfnamefont{M.}~\bibnamefont{Ernzerhof}},
  \bibinfo{journal}{Phys. Rev. Lett.} \textbf{\bibinfo{volume}{77}},
  \bibinfo{pages}{3865} (\bibinfo{year}{1996}).

\bibitem[{\citenamefont{Gyorffy}(1972)}]{Gyorffy1972}
\bibinfo{author}{\bibfnamefont{B.~L.} \bibnamefont{Gyorffy}},
  \bibinfo{journal}{Phys. Rev. B} \textbf{\bibinfo{volume}{5}},
  \bibinfo{pages}{2382} (\bibinfo{year}{1972}).

\bibitem[{\citenamefont{Liechtenstein et~al.}(1987)\citenamefont{Liechtenstein,
  Katsnelson, Antropov, and Gubanov}}]{Liechtenstein1987}
\bibinfo{author}{\bibfnamefont{A.~I.} \bibnamefont{Liechtenstein}},
  \bibinfo{author}{\bibfnamefont{M.~I.} \bibnamefont{Katsnelson}},
  \bibinfo{author}{\bibfnamefont{V.~P.} \bibnamefont{Antropov}},
  \bibnamefont{and} \bibinfo{author}{\bibfnamefont{V.~A.}
  \bibnamefont{Gubanov}}, \bibinfo{journal}{Journal of Magnetism and Magnetic
  Materials} \textbf{\bibinfo{volume}{67}}, \bibinfo{pages}{65 }
  (\bibinfo{year}{1987}).

\bibitem[{\citenamefont{Zener}(1951)}]{Zener1951}
\bibinfo{author}{\bibfnamefont{C.}~\bibnamefont{Zener}},
  \bibinfo{journal}{Phys. Rev.} \textbf{\bibinfo{volume}{81}},
  \bibinfo{pages}{440} (\bibinfo{year}{1951}).

\bibitem[{\citenamefont{Metropolis et~al.}(1953)\citenamefont{Metropolis,
  Rosenbluth, Rosenbluth, and Teller}}]{Metropolis1953}
\bibinfo{author}{\bibfnamefont{N.}~\bibnamefont{Metropolis}},
  \bibinfo{author}{\bibfnamefont{A.~W.} \bibnamefont{Rosenbluth}},
  \bibinfo{author}{\bibfnamefont{M.~N.} \bibnamefont{Rosenbluth}},
  \bibnamefont{and} \bibinfo{author}{\bibfnamefont{E.}~\bibnamefont{Teller}},
  \bibinfo{journal}{J.~Chem.~Phys.} \textbf{\bibinfo{volume}{21}},
  \bibinfo{pages}{1087} (\bibinfo{year}{1953}).

\bibitem[{\citenamefont{Binder}(1997)}]{Binder1997}
\bibinfo{author}{\bibfnamefont{K.}~\bibnamefont{Binder}},
  \bibinfo{journal}{Rep.~Prog.~Phys} \textbf{\bibinfo{volume}{60}},
  \bibinfo{pages}{487} (\bibinfo{year}{1997}).

\bibitem[{\citenamefont{Fischer et~al.}(2009)\citenamefont{Fischer, Dane,
  Ernst, Bruno, Lueders, Szotek, Temmerman, and Hergert}}]{Fischer2009}
\bibinfo{author}{\bibfnamefont{G.}~\bibnamefont{Fischer}},
  \bibinfo{author}{\bibfnamefont{M.}~\bibnamefont{Dane}},
  \bibinfo{author}{\bibfnamefont{A.}~\bibnamefont{Ernst}},
  \bibinfo{author}{\bibfnamefont{P.}~\bibnamefont{Bruno}},
  \bibinfo{author}{\bibfnamefont{M.}~\bibnamefont{Lueders}},
  \bibinfo{author}{\bibfnamefont{Z.}~\bibnamefont{Szotek}},
  \bibinfo{author}{\bibfnamefont{W.}~\bibnamefont{Temmerman}},
  \bibnamefont{and} \bibinfo{author}{\bibfnamefont{W.}~\bibnamefont{Hergert}},
  \bibinfo{journal}{Physical Review B (Condensed Matter and Materials Physics)}
  \textbf{\bibinfo{volume}{80}}, \bibinfo{pages}{014408}
  (\bibinfo{year}{2009}).

\bibitem[{\citenamefont{Choi et~al.}(2005)\citenamefont{Choi, Lee, Kim, Choi,
  Choi, Song, and Cho}}]{Choi2005}
\bibinfo{author}{\bibfnamefont{J.}~\bibnamefont{Choi}},
  \bibinfo{author}{\bibfnamefont{H.-W.} \bibnamefont{Lee}},
  \bibinfo{author}{\bibfnamefont{B.-S.} \bibnamefont{Kim}},
  \bibinfo{author}{\bibfnamefont{S.}~\bibnamefont{Choi}},
  \bibinfo{author}{\bibfnamefont{J.}~\bibnamefont{Choi}},
  \bibinfo{author}{\bibfnamefont{J.~H.} \bibnamefont{Song}}, \bibnamefont{and}
  \bibinfo{author}{\bibfnamefont{S.}~\bibnamefont{Cho}},
  \bibinfo{journal}{Journal of Applied Physics} \textbf{\bibinfo{volume}{97}},
  \bibinfo{pages}{10D324} (\bibinfo{year}{2005}).

\bibitem[{\citenamefont{Choi et~al.}(2006)\citenamefont{Choi, Lee, Kim, Park,
  Choi, Hong, and Cho}}]{Choi2006}
\bibinfo{author}{\bibfnamefont{J.}~\bibnamefont{Choi}},
  \bibinfo{author}{\bibfnamefont{H.-W.} \bibnamefont{Lee}},
  \bibinfo{author}{\bibfnamefont{B.-S.} \bibnamefont{Kim}},
  \bibinfo{author}{\bibfnamefont{H.}~\bibnamefont{Park}},
  \bibinfo{author}{\bibfnamefont{S.}~\bibnamefont{Choi}},
  \bibinfo{author}{\bibfnamefont{S.}~\bibnamefont{Hong}}, \bibnamefont{and}
  \bibinfo{author}{\bibfnamefont{S.}~\bibnamefont{Cho}},
  \bibinfo{journal}{Journal of Magnetism and Magnetic Materials}
  \textbf{\bibinfo{volume}{304}}, \bibinfo{pages}{e164} (\bibinfo{year}{2006}).

\bibitem[{\citenamefont{Choi et~al.}(2012)\citenamefont{Choi, Jo, Lee, Lee, Jo,
  Kajino, Takabatake, Ko, Park, and Jung}}]{Choi2012}
\bibinfo{author}{\bibfnamefont{Y.~H.} \bibnamefont{Choi}},
  \bibinfo{author}{\bibfnamefont{N.~H.} \bibnamefont{Jo}},
  \bibinfo{author}{\bibfnamefont{K.~J.} \bibnamefont{Lee}},
  \bibinfo{author}{\bibfnamefont{H.~W.} \bibnamefont{Lee}},
  \bibinfo{author}{\bibfnamefont{Y.~H.} \bibnamefont{Jo}},
  \bibinfo{author}{\bibfnamefont{J.}~\bibnamefont{Kajino}},
  \bibinfo{author}{\bibfnamefont{T.}~\bibnamefont{Takabatake}},
  \bibinfo{author}{\bibfnamefont{K.-T.} \bibnamefont{Ko}},
  \bibinfo{author}{\bibfnamefont{J.-H.} \bibnamefont{Park}}, \bibnamefont{and}
  \bibinfo{author}{\bibfnamefont{M.~H.} \bibnamefont{Jung}},
  \bibinfo{journal}{Applied Physics Letters} \textbf{\bibinfo{volume}{101}},
  \bibinfo{pages}{152103} (\bibinfo{year}{2012}).

\bibitem[{\citenamefont{Otrokov et~al.}(2013)\citenamefont{Otrokov, Borisova,
  Chis, Vergniory, Eremeev, Kuznetsov, and Chulkov}}]{Otrokov2013}
\bibinfo{author}{\bibfnamefont{M.~M.} \bibnamefont{Otrokov}},
  \bibinfo{author}{\bibfnamefont{S.~D.} \bibnamefont{Borisova}},
  \bibinfo{author}{\bibfnamefont{V.}~\bibnamefont{Chis}},
  \bibinfo{author}{\bibfnamefont{M.~G.} \bibnamefont{Vergniory}},
  \bibinfo{author}{\bibfnamefont{S.~V.} \bibnamefont{Eremeev}},
  \bibinfo{author}{\bibfnamefont{V.~M.} \bibnamefont{Kuznetsov}},
  \bibnamefont{and} \bibinfo{author}{\bibfnamefont{E.~V.}
  \bibnamefont{Chulkov}}, \bibinfo{journal}{JETP Letters}
  \textbf{\bibinfo{volume}{96}}, \bibinfo{pages}{714} (\bibinfo{year}{2013}).

\end{thebibliography}
\end{document}